\documentstyle[twocolumn,eqsecnum,epsf,aps]{revtex}

\begin{document}
\draft
\title{Magnetization-controlled spin transport in DyAs/GaAs layers}
\author{J. M. Mao, M. A. Zudov\footnotemark, R. R. Du}
\address{Department of Physics, University of Utah, Salt Lake City, UT 84112}
\author{P. P. Lee, L. P. Sadwick, and R. J. Hwu}
\address{Department of Electrical Engineering, University of Utah, Salt Lake 
City, UT
84112}
\date{\today}
\maketitle
\footnotetext{Presently at Stanford Picosecond Free Electron Laser Center, Stanford University, Stanford, CA 94305} 
\begin{abstract}
Electrical transport properties of DyAs epitaxial layers grown on
GaAs have been investigated at various temperatures and magnetic fields up to $12T$. The measured longitudinal resistances show two distinct peaks at fields around $0.2$ and $2.5T$ which are believed to be related to the strong
spin-disorder scattering occurring at the phase transition boundaries induced
by external magnetic field. An empirical magnetic phase diagram is
deduced from the temperature dependent experiment, and the anisotropic
transport properties are also presented for various magnetic field directions with respect to the current flow.
\end{abstract}


\preprint{HEP/123-qed}

\narrowtext

With the development of molecular beam epitaxy (MBE) techniques, rare-earth
monoarsenides (RE-As) can now be grown on GaAs substrate with
high quality \cite{r1,r2}. Such magnetic semimetals host a number of interesting
electronic and magnetic phases \cite{r3,r4,r5,r6,r7} mostly derived from the 
rare-earth
elements. Integration of magnetic semimetal with semiconductor, {\em e.g.} GaAs,
yields a new type of low-dimensional quantum structures where both charge and spin transport are of interest. The magnetotransport properties of
these compounds, however, have been relatively unexplored with the exception
of GaAs/ErAs/GaAs \cite{r1,r3,r4,r5}, for which the properties of carrier,
magnetic phases, and electronic band structure have been studied in some
detail. Here we report the results for magnetization-controlled spin
transport in DyAs thin layers measured at low-temperatures and high 
magnetic fields.

Samples were grown by MBE on semi-insulating GaAs $(001)$
substrates \cite{r2}. DyAs layer was grown at temperature $\sim 500^{\circ }C$ on top of the $200nm$ undoped GaAs buffer layer, and then an undoped 
GaAs cap layer of about $20nm$ was grown subsequently on top of DyAs.
Three samples having DyAs layer thickness of $70$, $270$, and $600nm$ have
been characterized. The structural quality of the DyAs layers \cite{r2} is
comparable with that of the GaAs/ErAs/GaAs grown by MBE \cite{r1}. For the
electrical transport measurements we used a Hall bar geometry
(width $500 \mu m$) defined by standard optical lithography. Indium contacts were alloyed at
$\sim 400^{\circ }C$ for electrical connections to both the electrons and holes
in the DyAs. The transport experiment was performed from room temperature
down to $0.4K$ in a $^3He$ refrigerator, and with a magnetic field $H$ up
to $12T$; a standard low-frequency $(<20Hz)$ lock-in technique was employed
for measurements of the magnetotransport coefficients. To study the
anisotropic properties, the magnetic field was oriented either perpendicular
to the plane of DyAs, or in the plane and with a varying direction to
the current flow $J$.

\begin{figure}[b]
\centerline{\epsfxsize=3.3in\epsfbox{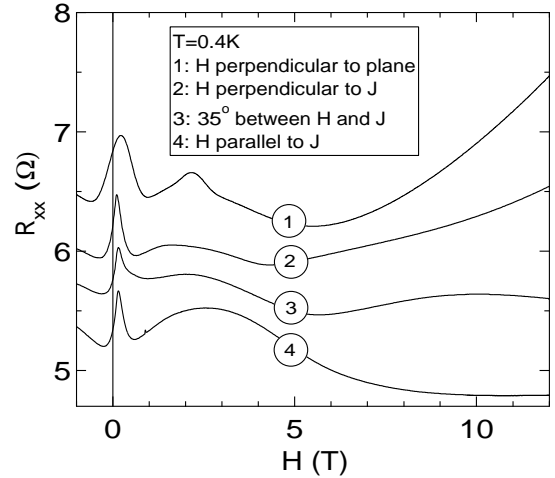}}
\caption{Magnetoresistances for the magnetic field perpendicular to the plane 
(line 1) and for three in-plane orientations with respect to the current flow 
(lines 2-4) at $0.4K$. The magnetic field was swept from negative to positive 
side, and the traces have been vertically offset for clarity.}
\label{fig1}
\end{figure}

Magnetoresistance characteristics are qualitatively similar for all
three samples. In the following, we concentrate on the results from the
sample with a DyAs layer thickness of $270nm$. In Fig.\ref{fig1}, we
present the longitudinal magnetoresistances, $R_{xx}$ , measured at four typical
magnetic field orientations at $0.4K$.
Except for the case where magnetic field
parallel to the current flow, positive background is seen in
magnetoresistance. In each of the $R_{xx}$ traces intrinsic signals from
the magnetization manifest themselves as two distinct peaks, seen here at
$\sim 0.2T$ and $\sim 2.5T$ respectively. 
We note that the $R_{xx}$ trace,
obtained here in a field sweep from negative to positive direction (from the left to the right in $H$ axis), is strongly asymmetric about $H=0$. We have reversed the field sweep direction and found that the curve is an exact mirror image of the previous trace. Such observations are unambiguous evidences for magnetization-controlled electronic transport in the DyAs layer.

Based on the discussion of the results in ErAs
\cite{r4}, we relate both of the peaks to the strong spin-disorder scattering occurring in magnetic phase transition regimes induced by external magnetic field. It should be mentioned that in GaAs/ErAs/GaAs only single magnetoresistivity peak has been observed at $\sim 1T$ and ascribed to antiferromagnet to paramagnet transition \cite{r4}. 
We tentatively
attribute the peak observed around $2.5T$ in DyAs to a transition of
similar nature. The origin of the anomaly at about $0.2T$ in our DyAs
samples is so far unresolved, but the resistance peak could be an indication
of the transition between two different configurations in the antiferromagnetic
phase. The possibility of multiple configurations in the antiferromagnetic
phase is consistent with the results from temperature-dependent
magnetization experiment, where two inflections showed up in the DyAs 
magnetization curves at temperatures around $6$ and $8K$, respectively \cite{r7}. Sharp peak and associated $R_{xx}$ change
influenced by weak magnetic field of $0.2T$ is, to some extent, similar to
giant magnetoresistance observed in metal superlattices and granular
materials \cite{r8}.

In contrast with the largely isotropic magnetization properties of DyAs
reported in \cite{r7}, the electrical transport properties are essentially anisotropic: $R_{xx}$ is very sensitive to the magnetic field-current flow configurations, {\em i.e.}, the peak amplitudes and their positions (especially for the peak around $2.5T$) vary with the field orientation. Furthermore, while it is absent in other field orientations, an additional peak in $R_{xx}$ emerges
at higher magnetic field of $\sim 9.5T$ for the in-plane magnetic field oriented
about $35^{\circ }$ with respect to the current flow. Strongly
anisotropic magnetoresistance in our experiments further indicates the
effects of crystal field interactions at low temperature. Such strong
effects were suggested by Child {\em et al} in their study on magnetic
properties of a variety of RE-V compounds by neutron diffractions \cite{r6}.
Like in other RE-V compounds \cite{r6}, the magnetically aligned 
sheets in DyAs are expected to be perpendicular to $(111)$ direction.
Magnetotransport experiments on DyAs grown on $(111)$ GaAs substrates, in addition to the present $(001)$ data, are thus needed to clarify the issue.

The Hall resistance $R_{xy}$ measured for the magnetic field perpendicular
to the DyAs layer is extremely small, of the order of $10^{-3}\Omega $.
Moreover, the overall shape of the Hall resistance is similar to that of $R_{xx}$, 
which may be caused by mixing of transport coefficients. Extremely 
small Hall resistance is due to either high carrier concentration or electron-hole compensation. Lack of the information on carrier density 
and mobility prevents us from quantitative analysis of the experimental
results. In particular, we were unable to assess the relative contribution from 
the electrons and that from the holes to the transport data, an issue demanding
further studies.
\begin{figure}[t]
\centerline{\epsfxsize=3.3in\epsfbox{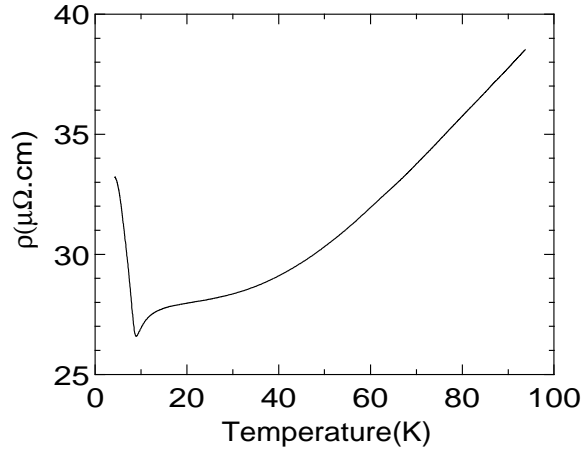}}
\caption{Longitudinal resistivity against temperature before magnetization.}
\label{fig2}
\end{figure}
As shown in Fig.\ref{fig2}, the temperature dependence of the resistivity 
before magnetization $(H=0)$ displays a dip around $8.5K$ followed by a sharp
increase at lower temperature, with a tendency to saturate below $4.4K$.
At the transition point from antiferromagnetism to paramagnetism,
a divergence of the resistivity's temperature derivative is expected, as
described in Reference \cite{r9}. The Neel temperature, $T_N$, could then be 
inferred from the maximum of $dR_{xx}(T)/dT$.
The estimated value $T_N \simeq 8K$ is consistent with the magnetization
measurements \cite{r7}.
\begin{figure}[b]
\centerline{\epsfxsize=3.5in\epsfbox{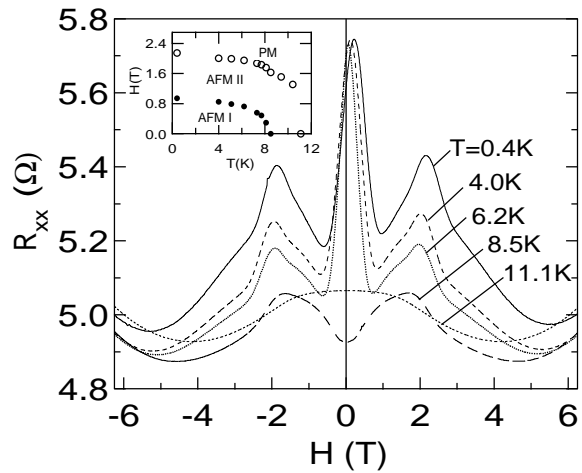}}
\caption{Longitudinal magnetoresistances for a magnetic field perpendicular to 
the plane at different temperatures. Inset: Magnetic field positions of valley (dots) and 
second peak (circles) in the longitudinal magnetoresistances at different 
temperatures, and the three magnetic phases (antiferromagnet I, II, and 
paramagnet) divided by them.}
\label{fig3}
\end{figure}
The $R_{xx}$ traces have been recorded at different temperatures; several 
typical curves are shown in Fig.\ref{fig3} for the $H$ field perpendicular to 
the plane and swept from negative to positive side. 
Again, the asymmetry of the 
$R_{xx}$ around $H=0$, which shows a strong $T$ dependence and
eventually disappears above $8K$, reflects the intrinsic magnetization in
the sample. It can also be seen from Fig.\ref{fig3} that both of the peak 
positions shift towards lower magnetic field as the temperature increases.
The amplitude of the first peak remains nearly unchanged up to $6.5K$ while
that of the second peak decreases monotonously within this temperature
range. At temperatures above $6.5K$, a drastic decrease of the first
peak is detected; this peak disappears at $8.5K$, which coincides with the Neel 
temperature $T_N \simeq 8K$. At the same time, the second peak shifts
rapidly to the low-field side and then disappears at about $11K$. 

In order to summarize our observations, we assume that the valley between the
two peaks represents the transition from one type of antiferromagnetism
configuration (i.e., AFM I) to another (AFM II), and the second peak the
transition point from the antiferromagnetism to paramagnetism (PM), to
arrive at an empirical magnetic phase diagram, as sketched in the inset of
Fig.\ref{fig3}. This phase diagram shows schematically the magnetic phase 
transition boundaries as critical magnetic field against temperature.

The temperature dependence of the magnetoresistance shows different
behavior for the in-plain magnetic field. 
Here we consider the case for an in-plane magnetic field parallel to the current
flow (not shown). The overall tendency of the first peak is nearly the same
as that in Fig.\ref{fig3}; it also disappears around $8K$. However, in contrast 
with the monotonous shift to the low-field side (as shown in Fig.3), the second 
peak first shifts to higher field up to $8.5K$, then moves down to the low-field 
side, and eventually approaches zero field at $16.5K$, {\em i.e.} higher 
temperature is required to convert the antiferromagnetism to paramagnetism. 
The experimental results from a series of field-current configurations lead us 
to conclude that the AFM I configuration of the antiferromagnetism is weakly 
anisotropic and the AFM II is strongly anisotropic. Such anisotropy is believed 
to be caused by a combination of the strong crystal field and the
strain/dislocations produced in the sample, since there is as large as two percent lattice mismatch between
DyAs and GaAs. The details of the magnetization-controlled 
anisotropic transport will be published elsewhere.

In summary, we have studied the magnetotransport properties of the epitaxial 
DyAs layers grown on GaAs. It is shown from the longitudinal
magnetoresistance data that the electronic transport is controlled by the
magnetization. The Neel temperature $T_N$ deduced from temperature dependence
of $R_{xx}$ is about $8K$, and is consistent with the magnetization results
\cite{r7}. We have observed two distinct peaks in the magnetoresistances which are attributed to strong spin-disorder scattering at the
magnetic phase boundaries. Our data suggest that there exist more than
one type of antiferromagnetism configurations in DyAs grown on $(001)$ 
GaAs, which is qualitatively different from ErAs where only
one peak has been observed. Strongly anisotropic transport properties further
support the notion that crystal field interaction plays an important role in
the magnetism of epitaxial DyAs layer on GaAs. 

We thank S. J. Allen, Jr. for discussions. This work is partially supported
by NSF.

\end{document}